# Joint User Association, Power Control and Scheduling in Multi-Cell 5G Networks

Bilal Maaz, Kinda Khawam, Samir Tohme
University of Versailles
France

Samer Lahoud
University of Rennes I
Rennes, France

Jad Nasreddine
Rafik Hariri University
Damour, Lebanon

*Abstract*—The focus of this paper is targeted towards multi-cell 5G networks which are composed of HPNs (High Power Node) such as evolved NodeBs ( HPNs) that control signaling and system broadcasting information and of simplified LPNs (Low Power Node) co-existing in the same operating area and sharing the scare radio resources. Consequently, greater emphasis is given to Inter-Cell Interference Coordination (ICIC) based on multi-resource management techniques that take in particular user association into account. Beside user association to HPNs, this paper takes also power control and scheduling into consideration. This addressed problem is remained largely unsolved, mainly due to its non-convex nature that makes the global optimal solution difficult to obtain. We address the user association challenge according to the two broadly adopted approaches in wireless networks: the network-centric approach where user association is allocated efficiently in a centralized fashion; and the user-centric approach where distributed allocation prevails for reduced complexity. The scheduling and HPNs power allocation are solved in a centralized fashion, in order to reach an optimal solution of the joint optimization problem.

## I. INTRODUCTION

5G networks are currently facing a significant challenge in terms of signaling load. Compared to its predecessors, 5G results in a significantly higher signaling requirement per subscriber. While a portion of this new signaling is required for new services and new devices types, the majority of signaling burden is related to mobility and paging. This increase is in part due to architectural changes such as heterogeneous networks and greater node density.

Consequently, the motivation of 5G networks is to enhance the capabilities of HPNs (High Power Node such as Base Stations) and simplify LPNs (Low Power Node) through connecting to a *signal processing cloud* with high-speed optical fibers. For a simplified architecture, all control signaling and system broadcasting information are delivered by HPNs to UEs. This paper addresses the issue of user association to HPNs in 5G networks. Further, as multiple HPNs use the same radio resources in a given operating area, ensuing interference harms radio transmissions and degrades the performances. Hence, a certain degree of coordination between the HPNs belonging to the same BBU pool is required to minimize the interference level through power control. Accordingly, the aim of this paper is threefold:

1. UE association to HPNs:
   - Decide the UE association to the adequate HPN for the signaling plan; this decision is operated in a centralized fashion by the advanced cloud computing processing techniques in the BBU pool
   - Decide the UE association to the adequate HPN for the data plan; this decision is operated by UEs in a distributed fashion.
2. Interference mitigation among HPNs through power control:
   - The joint HPN/UE association and HPN power control is solved in an iterative fashion involving the following steps:
   -- Fixing the assignment of UEs, the power levels are updated by HPNs to coordinate in alleviating inter-cell interference, thereby improving the overall network utility.
   -- Fixing the HPNs power allocation, the assignment of UEs to each HPNs is again done by solving the resulting optimization problem for UE association.
3. Fair scheduling of resources among UEs.

We consider Orthogonal Frequency Division Multiple Access (OFDMA) as the multi-access scheme for the downlink of 5G networks. As the same Resource Block (RB) is used in neighboring cells, interference may occur and degrade the channel quality of serviced UEs. Hence, efficient Inter-Cell Interference Coordination (ICIC) techniques [1] are still considered among the key building blocks of 5G networks, in particular, ICIC through power control. Multi-Resource management based on joint power allocation, scheduling and UE association is a primary key feature to improve global performance.

In this paper, we propose a unified framework to study the interplay of UE association and power allocation, in conjunction with fair scheduling. For that, we strive to optimize a network utility function which ensures proportional fairness among all serviced UEs.

### A. Related Work

The joint UE association (or alternately HPNs selection), scheduling and power control is a relevant problem in many wireless communications systems. However, despite its importance, is has remained largely unsolved, mainly due to its non-convex and combinatorial nature that makes the global optimal solution difficult to obtain.

In OFDMA networks, several articles have addressed the subject of joint UE association and power control ([2]-[7]). An intuitive idea is to optimize UE association and power levels in an iterative fashion, as suggested in ([2]-[4]). In [3], the authors propose an iterative method for power control and UE association: the power control is modeled as a non-cooperative game while the UE association relies on a signaling-based heuristic. The work in [4] considers a pricing-based UE association scheme for heterogeneous networks and proposes a distributed price update strategy based on a coordinate descent algorithm in the dual domain. The proposed UE association scheme is incorporated with power control and beamforming respectively and solved iteratively. The work in ([5]-[7]) strives to obtain global optimality for the joint UE association and power control problem. In [5], the joint problem is addressed by using duality theory, but only for a relaxed version of the problem where the discrete constraints are eliminated. In [6], the optimal settings for the UE association and power control that maximize the weighted sum rate are obtained under certain restricted conditions for the case where the number of UEs and HPNs is the same. Finally, authors in [7] propose a novel algorithm based on Benders decomposition to solve the joint non-convex problem optimally.

### B. Our Contribution

In our work, we show that proportional fairness among UEs boils down to time fairness in section III-A. The power control is solved in a centralized approach in section III-C, where for a fixed UE association, the power levels are updated by computing the resulting non-convex optimization problem for power control; the latter is rendered convex through geometric transformation. Such a solution allows multiple cells to coordinate in alleviating inter-cell interference, improving the overall network utility. The UE association in section IV is solved according to both the network-centric approach and the user-centric approach.

In the network-centric approach, we address the UE association in a centralized fashion by computing the resulting optimization problem.

In the user-centric approach, the UE association scheme is represented as non-cooperative game. In our case, HPNs and UEs optimize their local parameters by using of signaling messages already present in networks. A distributed algorithm for the UE association scheme based on Best-response algorithm will be applied by UEs to attain the Nash Equilibriums (NEs) of the game.

We have recourse to an iterative optimization approach involving a centralized power control for a fixed UE association and UE assignment for a fixed power allocation. The above two steps will be iterated to reach a (local) optimal solution of the joint optimization problem. The centralized schemes are stable but are highly computational. In fact, they require a central controller that collects information from HPNs and UEs, optimizes parameters, and sends signaling messages back to HPNs and UEs which can be cumbersome.

We address this multifaceted challenge according to the two broadly adopted approaches in wireless networks to better assess the resulting network performances.

The remainder of this paper is organized as follows. Section II presents the network model. Our approach is put forward in Section III. The scheduling problem is presented in Section III-A, the joint UE association and power control is explained in section III-D and the centralized power control is detailed in section III-C. The User association approach is explained in section IV where the network-centric UE association is presented in Section IV-A, and the user-centric UE association approach is presented in Section IV-B. Section V discusses simulation setup and displays quantitative results along with the discussion. Section VI concludes the paper.

## II. THE NETWORK MODEL

Consider the downlink of an OFDMA Single Input Single Output (SISO) cellular network, the radio Resource Block (RB) is the smallest radio resource unit [8] that can be scheduled to a mobile user. In order to evaluate the maximum system performance, a permanent downlink traffic scenario is considered, and all RBs are assigned at each scheduling period to a given UE.

Mathematical notations, variables and parameters used within this paper are defined in Table 1.

TABLE I. MATHEMATICAL NOTATIONS, VARIABLES AND PARAMETERS IN THE DOCUMENT.

| | |
|---|---|
| $J$ | Set of HPNs. |
| $I$ | Total set of UEs. |
| $K$ | Set of RBs. |
| $K(j)$ | Set of RBs used by HPN $j$ |
| $G_{ijk}$ | Channel power gain of UE $i$ on RB $k$ associated to HPN $j$. |
| $\rho_{ijk}$ | SINR of user $i$ associated HPN $j$ served on RB $k$. |
| $\beta_{jk}$ | Interference impact of all HPNs among UEs of HPN $j$. |

| $N_0$ | Noise power. |
|---|---|
| $\pi_{jk}$ | downlink power devoted by HPN $j$ to RB $k$. |
| $\alpha_{ij}$ | The proportion of time that UE $i$ is scheduled on the downlink by HPN $j$. |
| $p_j^{max}$ | Maximum downlink transmission power per HPN. |
| $p^{min}$ | Minimum downlink transmission power per RB. |

The Signal-to-interference-plus-noise-ration (SINR) of UE $i$ associated to HPN $j$ and allocated RB $k$ is given by:

$$\rho_{ijk} = \frac{\pi_{jk} G_{ijk}}{N_0 + \sum_{j' \neq j} \pi_{j'k} G_{ij'k}}. \quad (1)$$

We assume that there is a mapping function $f(\ )$ that maps $\rho_{ijk}$ to its corresponding bit rate $r_{ijk}$ (bit/s) realized by UE $i$ associated with HPN $j$ served on RB $k$, i.e., $r_{ijk} = f(\rho_{ijk})$.

Conventional UE association basically uses the max-SINR rule, it is evident from a network utility maximization perspective that max-SINR is inappropriate as it may deprive bad channel quality UEs from accessing radio resources. Hence, in this paper, we consider the network utility maximization problem under proportional fairness and we privilege users' interest by using the proportional equity incarnated by the logarithmic function according to the work in [9]. To reach this objective we maximize $\sum_{i \in I} \log(r_i)$ where $r_i$ is the mean bit rate of any UE $i$ given by:

$$r_i = \sum_{j \in J} \theta_{ij} \alpha_{ij} \sum_{k \in K(j)} f(\rho_{ijk}), \quad (2)$$

with $\theta_{ij}$ is the association variable given by what follows:

$$\theta_{ij} = \begin{cases} 1 \ if \ UE \ i \ is \ associated \ with \ eNB \ j \\ 0 \ otherwise. \end{cases} \quad (3)$$

Hence, the joint multi-resource management based on power control, UE association and proportional fair scheduling is as follows:

$$\underset{\theta, \alpha, \pi}{\text{maximize}} \sum_{i \in I} \log \left( \sum_{j \in J} \theta_{ij} \alpha_{ij} \sum_{k \in K(j)} f(\rho_{ijk}) \right) \quad (4a)$$

subject to: $\quad \sum_{j \in J} \theta_{ij} = 1, \forall i \in I, \quad (4b)$

$$\sum_{i \in I(j)} \alpha_{ij} = 1, \forall j \in j, \quad (4c)$$

$$\sum_{k \in K(j)} \pi_{jk} \leq P_j^{max}, \forall j \in J, \quad (4d)$$

$$\theta_{ij} \in \{0,1\}, \forall i \in I, \forall j \in J, \quad (4e)$$

$$0 \leq \alpha_{ij} \leq 1, \forall i \in I, \forall j \in J, \quad (4f)$$

$$\pi_{jk} \geq P_j^{min}, \forall j \in J, \forall k \in K(j). \quad (4g)$$

Constraints (4c) ensure that a UE is served at most 100% of the time by a given HPN. Constraints (4d-4e) guarantee the maximum total power consumed per HPN and the minimum power allocated per RB respectively. The utility function in (4a) can be re-written as:

$$U = \sum_{i \in I} \log \left( \sum_{j \in J} \theta_{ij} \alpha_{ij} \sum_{k \in K(j)} f(\rho_{ijk}) \right) \quad (5a)$$

$$= \sum_{i \in I} \sum_{j \in J} \theta_{ij} \log \left( \alpha_{ij} \sum_{k \in K(j)} f(\rho_{ijk}) \right) \quad (5b)$$

$$= \sum_{i \in I} \sum_{j \in J} \theta_{ij} \log(\alpha_{ij}) + \sum_{i \in I} \sum_{j \in J} \theta_{ij} \log(r_{ij}) \quad (5c)$$

where $r_{ij} = \sum_{k \in K(j)} \rho_{ijk}$ represents the mean bit rate obtained by UE $i$ connected to HPN $j$. In this paper, we consider that the function $f(\ )$ is the identity function. Accordingly, the utility formulation is technology-agnostic: the mapping between the throughput and the SINR of each UE can be derived in respect to the appropriate coding and modulation scheme in wireless networks. Inevitably, improving this network utility amounts to improving the UE throughput.

### III. PROBLEM FORMULATION

We show that proportional fairness among UEs boils down to time fairness in Section III-A. The ensuing joint UE association and power control problem will be presented in Section III-B. The centralized power control problem will be described in Section III-C.

#### A. The Scheduling Problem

The utility function in (5) contains in its first term the per cell scheduling problem that we intend to solve in this section (by computing $\alpha_{ij}$ which is the percentage of time UE $i$ is served in HPN $j$).

Assuming that UE $i$ has chosen HPN $l$ (i.e. $\theta_{il} = 1$; $\theta_{ij} = 0, \forall j \neq l$ ), we have what follows:

$$\sum_{i \in I, j \in J} \theta_{ij} \log(\alpha_{ij}) = \sum_{i \in I(l)} \log(\alpha_{il})$$

where $I(l)$ is the set of UEs associated to HPN $l$. Consequently, the scheduling problem for HPN $l$ is as follows:

$$\underset{\alpha}{\text{maximize}} \quad \sum_{i \in I(l)} \log(\alpha_{il}) \quad (6a)$$

subject to $\quad \sum_{i \in I(l)} \alpha_{il} = 1, \forall l \in j, \quad (6b)$

$$0 \leq \alpha_{il} \leq 1, \forall i \in I(l), \forall l \in J. \quad (6c)$$

*Proposition 3.1*: the optimal solution of the scheduling problem is given by what follows:

$$\alpha_{il}^* = \frac{1}{|I(l)|}, \forall l \in J, \forall i \in I(l). \quad (7)$$

**Proof:** Problem (6) is a convex optimization as the utility function (6a) is concave (sum of concave functions) and all constraints are linear. Let us express the KKT conditions that provide a first-order optimality condition for the problem:

$$-\frac{1}{\alpha_{il}} + \mu_l = 0, \quad \forall l \in J, \forall i \in I(l) \quad (8a)$$

$$\sum_{i \in I(l)} \alpha_{il} \leq 1, \quad \forall l \in J \quad (8b)$$

$$\mu_l \left( \sum_{i \in I(l)} \alpha_{il} - 1 \right) = 0, \quad \forall l \in J. \quad (8c)$$

From constraints (8a), we know that $\mu_l \neq 0$, otherwise $\frac{1}{\alpha_{il}} = 0$ which is not possible. Hence, we deduce from constraint (8c) that $\sum_{i \in I(l)} \alpha_{il} = 1$. Furthermore, the utility function in (6a) can be re-written as:

$$\log \left( \prod_{i \in I(l)} \alpha_{il} \right) \quad (9)$$

As the sum of the $\alpha_{il}$ variables is constant, the product of these variables is maximized for $\alpha_{il} = \frac{1}{|I(l)|}, \forall l \in J, \forall i \in I(l)$.

### B. The Joint UE Association and Power Control Problem

As $\alpha_{ij} = \frac{1}{|I(j)|} = \frac{1}{\sum_{i \in I(j)} \theta_{ij}}, \forall j \in J$, the utility function in (5) can be re-written such as:

$$U = \sum_{i \in I} \log \left( \sum_{j \in J} \frac{\theta_{ij}}{\sum_{i \in I(j)} \theta_{i'j}} \sum_{k \in K(j)} \rho_{ijk} \right) \quad (10)$$

As the $\theta_{ij}$ variables are binary and $\sum_{j \in J} \theta_{ij} = 1$ for all UEs, there exists only one HPN $j$ for which $\theta_{ij} = 1$ ($\theta_{ij'} = 0, \forall j' \neq j \in J$). Hence, the utility function in (10) can be re-casted as:

$$U = \sum_{i \in I} \sum_{j \in J} \theta_{ij} \log \left( \frac{\sum_{k \in K(j)} \rho_{ijk}}{1 + \sum_{i' \neq i} \theta_{i'j}} \right) \quad (11)$$

Given Jensen's inequality and the concavity of the log function, we have:

$$\log \left( \frac{\sum_{k \in K(j)} \frac{\rho_{ijk}}{1 + \sum_{i' \neq i} \theta_{i'j}}}{|K(j)|} \right) \geq \frac{\sum_{k \in K(j)} \log \left( \frac{\rho_{ijk}}{1 + \sum_{i' \neq i} \theta_{i'j}} \right)}{|K(j)|}$$

Thus, the utility function can be re-casted as follows:

$$U = \sum_{i \in I} \sum_{j \in J} \theta_{ij} \log \left( \sum_{k \in K(j)} \frac{\rho_{ijk}}{1 + \sum_{i' \neq i} \theta_{i'j}} \right) \quad (12)$$

$$\geq \sum_{i \in I} \sum_{j \in J} \sum_{k \in K(j)} \log \left( \frac{\rho_{ijk}}{1 + \sum_{i' \neq i} \theta_{i'j}} \right)$$

We denote by $\bar{U}$ the upper bound on the utility function, given by:

$$\bar{U} = \sum_{i \in I} \sum_{j \in J} \sum_{k \in K(j)} \log \left( \frac{\rho_{ijk}}{1 + \sum_{i' \neq i} \theta_{i'j}} \right) \quad (13)$$

Henceforward, we adopt this newly defined utility function $\bar{U}$. The ensuing joint UE association and power control will be solved according to a centralized power control in section III-C and two UE association approaches in section IV.

### C. The Centralized Power control

In this section, we resort to a centralized power control scheme by fixing the UE association, the corresponding Power Control (PC) problem is given by:

$$\text{maximize}_{\pi} U^{PC}(\pi) = \quad (14a)$$

$$\sum_{j \in J} \sum_{i \in I(j)} \sum_{k \in K(j)} \log \left( \frac{1}{n_j} \times \frac{\pi_{jk} G_{ijk}}{N_0 + \sum_{j' \neq j} \pi_{j'k} G_{ij'k}} \right)$$

subject to:
$$\sum_{k \in K(j)} \pi_{jk} \leq P_j^{max}, \forall j \in J, \quad (14b)$$

$$\pi_{jk} \geq P_j^{min}, \forall j \in J, \forall k \in K(j). \quad (14c)$$

where $n_j$ is the number of UEs associated to HPN j, i.e. |I(j)|.

Problem (14) in a non-convex optimization problem. However, it can be rendered convex through geometric programming by performing a variable change $\hat{\pi}_{jk} = \log(\pi_{jk})$ and defining the following $\hat{N}_0 = \log(N_0)$, $\hat{G}_{ijk} = \log(G_{ijk})$ and $\hat{n}_j = \log(n_j)$. The resulting optimization problem deemed $P(\hat{\pi})$ is given by the following:

$$\text{maximize}_{\pi} U^{PC}(\hat{\pi}) = \quad (15a)$$

$$\sum_{j \in J} \sum_{i \in I(j)} \sum_{k \in K(j)} (\hat{\pi}_{jk} + \hat{G}_{ijk} - \log(\exp(\hat{N}_0)$$

$$+ \sum_{j' \neq j} \exp(\hat{\pi}_{j'k} + \hat{G}_{ij'k}))) - \sum_{j \in J} K. n_j. \hat{n}_j$$

subject to:

$$\log \left( \sum_{k \in Kj)} \exp(\hat{\pi}_{jk}) \right) - \log(p_j^{max}) \leq 0. \quad (15b)$$

$$-\hat{\pi}_{jk} + \log(p_j^{min}) < 0, \forall j \in J, \forall k \in K(j). \quad (15c)$$

*Proposition 4.1:* The optimization problem $P(\hat{\pi})$ is convex.

**Proof:** The first part of the utility function is linear thus concave. The second part includes the log-sum-exp expressions which are convex and hence their opposite is concave. Further, the new constraints in (15b) are convex owing to the properties of the log-sum-exp expression, while the constraints in (15c) are linear and hence convex.

## IV. THE USER ASSOCIATION APPROACHES

In Addition to the centralized power control scheme presented in Section III-C, in this Section we present the UE association schemes. Both schemes will be run iteratively until convergence. For the network-centric approach, the convergence to a local optimum is guaranteed as, at each iteration, both the power control scheme and the UE association scheme monotonically improve the value of the utility function. For the user-centric approach we resort to a fully *distributed UE association scheme* based on Best-response algorithm will be run by UEs.

### A. Network-centric UE Association

For fixed power levels, The UE Association (UA) problem is given by:

$$\underset{\theta}{\text{maximize}} \; U^{UA}(\pi) = \sum_{i \in I} \sum_{j \in J} \sum_{k \in K} \theta_{ij} \log\left(\frac{\rho_{ijk}}{n_j}\right) \quad (16a)$$

Subject to:
$$\sum_{j \in J} \theta_{ij} = 1, \forall i \in I, \quad (16b)$$

$$\theta_{ij} \in \{0,1\}, \forall i \in I(j) \quad (16c)$$

$$n_j = \sum_{i \in I} \theta_{ij}, \forall j \in J. \quad (16d)$$

The problem in (16) is combinatorial due to the binary variable $\theta_{ij}, \forall j \in J, \forall i \in I$ and the complexity of the brute force algorithm (in $O(|J|^{|I|})$) is exponential in the number of UEs. A workaround is to allow UEs to be associated to more than one HPNs, i.e., the UE association becomes a load balancing scheme. The relaxed problem, that we deem *optimal load balancing*, is convex (for $0 \leq \theta_{ij} \leq 1$,) and provides an upper bound to the original problem in (16). However, in a practical system, it is much more difficult to implement a load balancing algorithm that allows costly recurrent shifts between HPNs than a UE association algorithm (single HPN selection). Thus, we adopt a rounding method to revert back to the original UE association problem and we deem it *centralized UE Association*.

### B. User-centric UE Association

We propose to solve the distributed UE association problem by having recourse to non-cooperative game theory. Non-Cooperative game theory models the interactions between players competing for a common resource. Hence, it is well adapted to model the HPN selection scheme. Here UEs are the decision makers or players of the game. We define a multiplayer game $G_{UA}$ between the |I| UEs, assumed to make their decisions without knowing the decisions of each other.

The formulation of this non-cooperative game $G_{UA} = \langle I, S, U^{UA} \rangle$ can be described as follows:

- A finite set of UEs i={1,...,|I|}.
- The space of pure strategies S formed by the Cartesian product of each set of pure strategies $S = S_1 \times S_2 \times ... \times S_{|I|}$, where the strategy space of any UE $i$ is $S_i = \{eNB_j^i, eNB_{j'}^i\}$ with $j, j' \in J$.
  - If the UE i is finally associated with $eNB_j^i$ (this is an outcome of the pure strategies played by UE i), then $\theta_{ij} = 1$, else $\theta_{ij'} = 1$.
  - We denote by $a_i \in S_i$ the action taken by UE i.
- A set of utility functions
$$U^{UA} = \left(U_1^{UA}(\theta), U_2^{UA}(\theta), ..., U_{|I|}^{UA}(\theta)\right)$$
That quantify UEs' utility for a given strategy profile θ, where the utility function of any UE i is given by:

$$U^{UA} = \sum_{j \in J} \sum_{k \in K} \theta_{ij} \log\left(\frac{\rho_{ijk}}{1 + \sum_{i' \neq i} \theta_{i'j}}\right) \quad (17)$$

The game $G_{UA}$ is an unweighted crowding game as it is a normal-form game in which the UEs share a common set of actions and the payoff a particular UE *i* receives for choosing a particular action (selecting one of the available HPNs) is player specific and a non-increasing function of the total number of UEs choosing that same action. Unweighted crowding games have PNE (Pure NE). Furthermore, when players have only two strategies (choosing between $eNB_j^i$ and $eNB_{j'}^i$ for any UE i), the game has the Finite Improvement Path (FIP) property [10] and hence a Best-Response algorithm permits attaining the PNE of the game. In fact, according to the optimal UE association as investigated in the performance evaluation Section VI, the large majority of UEs will be only associated to a single HPN and very few UEs will load balance their traffic among two HPNs solely. Hence, in the user-centric approach deem distributed approach, it is largely enough to give each UE a choice among the two strategies denoted $(\theta_i, 1 - \theta_i)$. Accordingly, the utility function in (17) can be re-written as:

$$U_i^{UA} = \quad (18)$$

$$\theta_i \log\left(\frac{\rho_{ij}}{1+\sum_{i'\neq i}\theta_{i'}}\right) + (1-\theta_i) \log\left(\frac{\rho_{ij'}}{|I|-\sum_{i'\neq i}\theta_{i'}}\right) =$$

$$\theta_i \log\left(\frac{\rho_{ij}/(1+\sum_{i'\neq i}\theta_{i'})}{\rho_{ij'}/(|I|-\sum_{i'\neq i}\theta_{i'})}\right) + \log\left(\frac{\rho_{ij'}}{|I|-\sum_{i'\neq i}\theta_{i'}}\right)$$

where $\rho_{ij} = \prod_{k\in K}\rho_{ijk}$. Note that the second term in (18) is independent of the player strategy and does not intervene in the strategy updates given in algorithm 1. Further at each round of the Best-response algorithm, each UE $i$ favors the HPN that endows it with the higher mean rate according to (19).

**Algorithm 1** RL algorithm for UE Association

1) Initialization: set t=0 and each UE i defines an initial strategy $\theta_i(0)$.
2) For i={1,...,|I|}, do:
   - For each UE $i$, if

$$\frac{\rho_{ij}}{1+\sum_{i'\neq i}\theta_{i'}} > \frac{\rho_{ij'}}{|I|-\sum_{i'\neq i}\theta_{i'}} \quad (19)$$

   Then UE i associates with $eNB_j^i(\theta_i(t)=1)$.
   - Else, UE i associates with $eNB_{j'}^i(\theta_i(t)=0)$.

3) Set $t \leftarrow t+1$ and go to step 2 (until satisfying termination criterion: $\theta_i(t+1) = \theta_i(t)$ for all $i \in I$).

## V. PERFORMANCE EVALUATION

We consider a network with a bandwidth of 5 MHZ with 9 hexagonal cells. The physical layer parameters are based on 3GPP technical specifications TS 36.942 [11]. The simulation parameters are displayed in Table 2.

TABLE II. PHYSICAL LAYER AND SIMULATION PARAMETERS

| *Channel bandwidth (MHz)* | 5 | *Number of RBs* | 25 |
|---|---|---|---|
| *Thermal noise (dBm)* | -104.5 | *Time subframe TTI (ms)* | 1 |
| *Max power/eNodeB (dBm)* | 43 | *Min Power/RB (dBm)* | 15 |
| *Number UE/eNodeB* | 4 to 14 | *Number eNodeBs* | 9 |
| *Frequency reuse* | 1 | *User noise figure (dB)* | 7.0 |
| *Antenna configuration* | | 1-transmit, 1-receive SISO (Single Input Single Output) | |

In this paper, we conducted preliminary simulations in a Matlab simulator, where various scenarios were tested to assess the performances of our control schemes. For each approach, 25 simulations were run where in each cell a predefined number of users is selected; users' positions were uniformly distributed in the cells.